\documentclass[sigconf,nonacm]{acmart}

\usepackage[most]{tcolorbox}
\usepackage{empheq}
  
\newtcbox{\mymath}[1][]{%
    nobeforeafter, math upper, tcbox raise base,
    enhanced, colframe=blue!30!black,
    colback=blue!30, boxrule=1pt,
    #1}
    
\newcommand{\degree}{\mathrm{deg}}
\usepackage{amsmath}
\usepackage{tikz}
\usepackage{mathdots}

\usepackage{amsfonts}
\usepackage{amsthm}

\usepackage{amssymb}
\usepackage{latexsym}   
\usepackage{graphicx}
\usepackage{color}
\usepackage{hyperref}
\usepackage[linesnumbered,ruled,vlined]{algorithm2e}
\usepackage{booktabs}
\usepackage{arydshln}
\usepackage{bm}
\usepackage{multirow}
\usepackage{makecell}
\usepackage{booktabs}
\usepackage{balance}
\usepackage{subcaption}

\newcommand{\argmax}{\mathop{\rm argmax}}
\newcommand{\argmin}{\mathop{\rm argmin}}

\newcommand{\bp}{\bm{p}}
\newcommand{\bq}{\bm{q}}
\newcommand{\bbR}{\mathbb{R}}

\newcommand{\hide}[1]{}

\newtheorem{problem}{Problem}

\def\methodlag{\texttt{Q-Lagrange}\xspace}
\def\method{\texttt{Q-Peeling}\xspace}
\def\myproblem{\textsf{Q-DISCO}\xspace}

\AtBeginDocument{%
  }

\copyrightyear{2025}
\acmYear{2025}
\setcopyright{acmlicensed}\acmConference[WSDM '25]{Proceedings of the Eighteenth ACM International Conference on Web Search and Data Mining}{March 10--14, 2025}{Hannover, Germany}
\acmBooktitle{Proceedings of the Eighteenth ACM International Conference on Web Search and Data Mining (WSDM '25), March 10--14, 2025, Hannover, Germany}
\acmDOI{10.1145/3701551.3703502}
\acmISBN{979-8-4007-1329-3/25/03}

\begin{document}

\title[Q-DISCO: Query-Centric Densest Subgraphs in Networks with Opinion Information]{Q-DISCO: Query-Centric Densest Subgraphs in Networks\\ with Opinion Information}

 \author{Tianyi Chen}
 \email{t.chen@celonis.com}
 \affiliation{%
   \institution{Celonis}
   \city{New York}
   \state{NY}
   \country{USA}
 }
 
 \author{Atsushi Miyauchi}
 \email{atsushi.miyauchi@centai.eu}
 \orcid{0000-0002-6033-6433}
 \affiliation{%
   \institution{CENTAI Institute}
   \city{Turin}
   \country{Italy}
 }
 
 \author{Charalampos E. Tsourakakis}
 \email{charalampos.tsourakakis@relational.ai}
 \affiliation{%
   \institution{RelationalAI}
   \city{Boston}
   \state{MA}
   \country{USA}
 }

\renewcommand{\shortauthors}{Chen, Miyauchi, Tsourakakis}

\begin{abstract}
Given a network $G=(V,E)$, where each node $v$ is associated with a vector $\bp_v \in \mathbb{R}^d$ representing its opinion about $d$ different topics, how can we uncover subsets of nodes that not only exhibit exceptionally high density but also possess positively aligned opinions on multiple topics?  In this paper we focus on this novel algorithmic question, that is essential in an era where digital social networks are hotbeds of opinion formation and dissemination.
We introduce a novel methodology anchored in the well-established densest subgraph problem. We analyze the computational complexity of our formulation, indicating that our problem is NP-hard and eludes practically acceptable approximation guarantees. To navigate these challenges, we   design two heuristic algorithms: the first is predicated on the Lagrangian relaxation of our formulation, while the second adopts a peeling algorithm based on the dual of a Linear Programming relaxation. We elucidate the theoretical underpinnings of their performance and validate their utility through empirical evaluation on  real-world datasets.  Among others, we delve into Twitter datasets we collected concerning timely issues, such as the Ukraine conflict and the discourse surrounding COVID-19 mRNA vaccines, to gauge the effectiveness of our methodology. Our empirical investigations verify that our algorithms are able to extract valuable insights from networks with opinion information.

\end{abstract}




\maketitle

\section{Introduction}
\label{sec:intro} 
Advances in Artificial Intelligence (AI), particularly with the advent of large language models~\cite{brown2020language,devlin2018bert}, coupled with the ubiquity of social media platforms~\cite{kwak2010twitter,kaplan2010users}, have paved the way for unparalleled opportunities in capturing and understanding the vast and complex networks of human opinions. These sophisticated AI technologies can analyze and synthesize vast amounts of data generated on social media, enabling the identification of patterns, trends, and sentiments across diverse populations. This convergence of AI and social media not only enhances our ability to map out the intricate web of opinions that shape public discourse but also offers a more nuanced and accurate representation of societal views and ideologies. This necessitates the development of novel algorithmic tools capable of answering complex queries that can potentially foster  more informed decision-making and facilitate a more connected and understanding global community.

This research aims to address a critical question in the analysis of complex networks: how can we identify a densely-connected group of nodes whose opinions across various topics closely align with a specified query vector? This question is pivotal for exploring how ideas or preferences cluster within networks, offering insights into the formation of consensus, the dynamics of social influence, and the structure of collective opinions. By focusing on this question,  we are able to detect cohesive communities with the same basic stance across different topics. Additionally, this approach is particularly valuable in collaborative filtering~\cite{Koren+22} and team formation~\cite{lappas2009finding}, where a new node with a given opinion vector (e.g., a user of a music subscription service that enjoys \textsf{Pop} and \textsf{Rock} but dislikes \textsf{Rap} and \textsf{Trap}) seeks a recommendation for a compatible community. Furthermore, in political science,   understanding the degree of agreement or polarization within groups is crucial~\cite{mehlhaff2022group,dandekar2013biased,musco2018minimizing,chen2022polarizing}. An algorithmic tool answering the aforementioned question allows for the examination of how closely knit a group is in terms of shared opinions and can help in strategies aimed at consensus building or in studying the effects of polarization.  

\smallskip
\noindent
\textbf{Our contribution.} 
We summarize our contributions as follows:

\begin{itemize}
\leftskip=-15pt
\item We introduce \myproblem, a novel optimization formulation for finding a query-centric densest subgraph in networks with opinion information. We analyze its hardness and prove that it is NP-hard in general and at least as hard to approximate as the Densest $k$-Subgraph (D$k$S) up to a constant factor, implying that we cannot expect any polynomial-time algorithm that possesses a practically acceptable approximation ratio. 

\item In the light of the hardness results, we design two heuristic algorithms with solid mathematical foundations. The first one called \methodlag is based on the Lagrangian relaxation, while the second one called \method adopts the well-known greedy peeling algorithm based on the dual of a Linear Programming (LP) relaxation.   We elucidate the
theoretical underpinnings of their performance, providing a-posteriori approximation guarantees. 

\item We carry out extensive experiments with real-world datasets, showing that our proposed methodology performs exceptionally well in a range of practical situations. Our findings are illustrated through three case studies: (i) identifying node subsets that reveal the intricate patterns of opinion dynamics in Twitter data, (ii) distinguishing overlapping dense clusters within the DBLP dataset in the Computer Science field, and (iii) deriving insights beneficial for collaborative filtering from Deezer data.  Our code is accessible to the public at \url{https://github.com/tsourakakis-lab/q-disco}.

\end{itemize}

\section{Related work}
\label{sec:related} 

Dense Subgraph Discovery (DSD) is a fundamental graph mining area whose goal is to find dense components in a given network~\cite{aggarwal,Gionis_Tsourakakis_15,tsourakakis2013denser}. 
Among many optimization models for DSD, the most well-studied is the Densest Subgraph Problem (DSP)~\cite{Lanciano+23}. 
In the most basic version, given an undirected graph $G=(V,E)$, consisting of $n=|V|$ nodes and $m=|E|$ edges, we are asked to find a node subset $S\subseteq V$ that maximizes the so-called \emph{density} $d(S)=|E(S)|/|S|$, where $E(S)$ is the set of edges in the subgraph $G[S]$ induced by $S$. 
The problem is known to be polynomial-time solvable using $O(\log n)$ max-flow computations~\cite{Goldberg84,Picard82}, $O(1)$ number of parametric flows~\cite{gallo1989fast},  or LP~\cite{Charikar00}. 
Owing to its intuitive definition and solvability, DSP has actively been extended to various settings, including directed graphs~\cite{ma2020efficient}, dynamic graphs~\cite{Bhattacharya+15}, streaming graphs~\cite{esfandiari2015applications,bahmani2012densest,mcgregor2015densest}, community detection~\cite{Miyauchi_Kakimura_18}, uncertain scenarios~\cite{Miyauchi_Takeda_18,Kuroki+20}, negative edge weights~\cite{tsourakakis2020novel}, graph decomposition~\cite{danisch2017large,balalau2015finding,veldt2021generalized}, higher-order extensions ~\cite{Huang+95,tsourakakis2015k,mitzenmacher2015scalable,chen2022algorithmic,sun2020kclist++}, and multilayer networks~\cite{Jethava+15,Galimberti+17,Kawase+23}. 
For DSP, there also exists a linear-time $1/2$-approximation algorithm, which greedily removes a node with the smallest degree and reports the node subset achieving the maximum density seen among those subsets~\cite{Kortsarz_Nutov_05,Charikar00}. 
 This kind of algorithm is often called the greedy peeling algorithm. 
Recently, Boob et al.~\cite{boob2020flowless}, inspired by the Multiplicative Weights Update method~\cite{arora2012multiplicative}, proposed {\sc Greedy++}, an iterative peeling algorithm that generalizes the above, and empirically demonstrated that the algorithm converges to a near-optimal solution extremely fast. Very recently, Chekuri et al.~\cite{chekuri2022densest} theoretically proved that {\sc Greedy++} indeed converges to the optimum, in a wider problem setting.

The problem called the Densest $k$-Subgraph (D$k$S), a key challenge within size-constrained DSP, requires identifying $S\subseteq V$ of size $k$ that maximizes $d(S)$ or equivalently $|E(S)|$, and is known to be NP-hard. The current best approximation algorithm for D$k$S offers an $\Omega(1/n^{1/4+\epsilon})$-approximate solution in $n^{O(1/\epsilon)}$ time for any $\epsilon >0$~\cite{Bhaskara+10}, while it is also known that D$k$S cannot be approximated within a factor of $n^{\frac{1}{(\log\log n)^c}}$ for some $c>0$ under the Exponential Time Hypothesis (ETH). Andersen and Chellapilla~\cite{Andersen_Chellapilla_09} introduced two variations of D$k$S: the Densest at-Least-$k$-Subgraph (Dal$k$S) and Densest at-most-$k$-Subgraph (Dam$k$S) problems, with the former requiring $|S|\geq k$ and the latter $|S|\leq k$. While Dal$k$S has a $1/3$-approximation via a greedy peeling algorithm~\cite{Andersen_Chellapilla_09}, Khuller and Saha~\cite{Khuller_Saha_09} improved this to a $1/2$-approximation using an LP-based approach and also established its NP-hardness. They further demonstrated that approximating Dam$k$S is at least as hard as approximating D$k$S, up to a constant factor. Our formulation, \myproblem, extends Dal$k$S to a node-weighted setting, accommodating possibly negative weights and also encompasses Dam$k$S, offering a new perspective on these problems (see Section~\ref{sec:problem}).

Analyzing social networks with opinion information plays a key role in social network analysis~\cite{Bakshy+12}. The field of opinion dynamics has attracted considerable attention from researchers in various disciplines, as underscored by the tutorials from Proskurnikov and Tempo~\cite{proskurnikov2017tutorial,proskurnikov2018tutorial}. Additionally, a substantial body of recent research has concentrated on the model-driven optimization of societal objectives, such as reducing polarization~\cite{gionis2013opinion, gaitonde2020adversarial, abebe2021opinion, musco2018minimizing, ristache2024wiser}. However, there is a notable scarcity in the development of graph mining primitives designed for graphs that incorporate opinion information, a line of research that our work seeks to advance.
In the context of cohesive subgraph detection, Anagnostopoulos et al.~\cite{anagnostopoulos2020spectral} and Miyauchi et al.~\cite{Miyauchi+23} studied node-attributed graphs and aimed to extract a densest subgraph that is fair (or more generally, diverse) in terms of the attributes. 
When dealing with discrete opinions (such as neutral, positive, or negative stances on a topic), these primitives can be employed to derive cohesive subgraphs that avoid the over-representation of any specific opinion. 
The work most closely related to ours is that of Fazzone et al.~\cite{fazzone2022discovering}, who introduced the Heavy and Dense Subgraph Problem (HDSP), which asks to maximize the generalized density formula incorporating nonnegative node weights $\bm{w}=(w_v)_{v\in V}$, i.e., \((|E(S)|+\sum_{v\in S}w_v)/|S|\).
Very recently, Huang et al.~\cite{huang2023densest} addressed a generalization of HDSP, where node weights are potentially negative, which we call HDSP with Positive and Negative node weights (HDSP-PN). Remarkably, Huang et al.~\cite{huang2023densest} developed an algorithm that solves HDSP-PN in strongly-polynomial time (i.e., $O(n)$ max-flow computations). Our model is a variant of HDSP-PN, applying a lower bound constraint on average node weight, rather than maximizing this metric alongside density, enabling us to control the trade-off between the density and the affinity of opinions by varying the threshold.

\section{Problem Formulation and Hardness}\label{sec:problem}

In this section, we first mathematically formulate our problem. Then we prove the NP-hardness and some hardness of approximation, based on those of D$k$S and Dam$k$S. 

\subsection{Problem formulation}
Let $G=(V,E)$ be an undirected graph consisting of $n=|V|$ nodes and $m=|E|$ edges. 
Each node $v\in V$ has an opinion vector $\bp_v\in \bbR^d$ on $d$ different topics. 
Given a query opinion vector $\bq\in \bbR^d$, we define the \textit{agreement} of node $v\in V$ to $\bq$ as the dot product between the opinion vector of $v$ and the query vector, 
denoted by $c_v =\bp_v \cdot \bq$. 
Note that $c_v$ can be negative, representing the disagreement between them. 
For any $S\subseteq V$, we also define the agreement of $S$ to $\bq$ as the average value of the agreements over $v\in S$, denoted by $c(S)=\frac{\sum_{v\in S}c_v}{|S|}$. 
Our problem aims to find a node subset $S\subseteq V$ that is as dense as possible while satisfying the constraint that the agreement of $S$ to the given query vector $\bq$ is no less than the given threshold $\theta$. 
Formally, the problem can be formulated as follows: 
\begin{problem}[\myproblem]
\label{prob:qdisco}
Given an undirected graph $G=(V,E)$, where each node $v\in V$ has an opinion vector $\bp_v\in \mathbb{R}^d$ on $d$ different topics, a query vector $\bq\in \mathbb{R}^d$, and a threshold $\theta\in \mathbb{R}$, we are asked to find $S\subseteq V$ that maximizes the density $d(S)$ subject to the constraint that the agreement of $S$ to $\bq$ is greater than or equal to $\theta$, i.e., $c(S)\geq \theta$. 
\end{problem}

This problem can be seen as a constrained variant of HDSP-PN, studied by Huang et al.~\cite{huang2023densest}. 
Indeed, HDSP-PN aims to maximize the density and the average node weight simultaneously, 
while our problem maximizes only the density but imposes the lower bound on the average node weight, 
enabling us to control the trade-off between the density and the average node weight (i.e., affinity of opinions) by varying the threshold $\theta$. 
Conversely, HDSP-PN can be seen as a relaxation of our problem. Specifically, a Lagrangian relaxation of our problem becomes an instance of HDSP-PN.

\subsection{Hardness}

Unlike HDSP-PN being polynomial-time solvable~\cite{huang2023densest}, our problem is NP-hard, even for quite limited instances: 
\begin{proposition}\label{prop:reduction}
\myproblem is NP-hard, even for the instances in which $c_v=0$ or $1$ for every $v\in V$. 
\end{proposition}

The proof can be found in Appendix~\ref{apdx:prop1}. Note that the proof gives a polynomial-time approximation-preserving reduction from Dam$k$S to \myproblem only with the instances in which $c_v=0$ or $1$ for every $v\in V$. 
Recall that if there exists a polynomial-time $\alpha$-approximation algorithm for Dam$k$S, then there exists a polynomial-time $\alpha/4$-approximation algorithm for D$k$S~\cite{Khuller_Saha_09}. 
Therefore, we have the following result: 
\begin{proposition}\label{prop:approx}
If there exists a polynomial-time $\alpha$-approximation algorithm for \myproblem (only with the instances in which $c_v=0$ or $1$ for every $v\in V$), there exist polynomial-time $\alpha$-approximation and $\alpha/4$-approximation algorithms for Dam$k$S and D$k$S, respectively. 
\end{proposition}

Combining the above with the fact that assuming the Exponential Time Hypothesis (ETH), D$k$S cannot be approximated up to a factor of $n^\frac{1}{(\log\log n)^c}$ for some $c>0$~\cite{Manurangsi+17}, we have the following result: 
\begin{proposition}\label{prop:eth_hardness}
Assuming ETH, \myproblem (only with the instances in which $c_v=0$ or $1$ for every $v\in V$) cannot be approximated up to a factor of $n^\frac{1}{(\log\log n)^c}$ for some $c>0$. 
\end{proposition}

Therefore, we cannot expect any polynomial-time algorithm for \myproblem that possesses a practically acceptable approximation ratio, which motivates us to design effective heuristic algorithms.

\section{The Proposed Methods} 
In this section, we design two heuristic algorithms, both of which have solid mathematical foundations. The first one is based on a Lagrangian relaxation of the problem, while the second one adopts a greedy peeling algorithm based on the dual of an LP relaxation. 
Throughout this section, we assume without loss of generality that $\theta < \max_{v\in V}c_v$ holds and there exist $u,v\in V$ such that $c_u\neq c_v$ (i.e., otherwise the problem simply reduces to the DSP). 

\subsection{Lagrangian relaxation-based algorithm}\label{subsec:Lagrangian}

Our proposed algorithm ($\methodlag$) is based on a Lagrangian relaxation~\cite{boyd2004convex} of \myproblem, which is defined using a Lagrangian multiplier $\lambda\geq 0$ as follows:
\begin{align*}
&\max_{S\subseteq V}\, H_\lambda(S) := d(S) + \lambda (c(S) - \theta). 
\end{align*}
Notice that the above Lagrangian relaxation is an instance of HDSP-PN, studied by Huang et al.~\cite{huang2023densest}. 
Indeed, it suffices to consider the instance $(G,\bm{w})$ of HDSP-PN, where $\bm{w}=(w_v)_{v\in V}$ such that $w_v=\lambda\cdot c_v$ for $v\in V$. 
Using the algorithm by Huang et al.~\cite{huang2023densest}, we can solve the Lagrangian relaxation in strongly-polynomial time. 

Varying $\lambda\geq 0$, we can see the spectrum of the strength of the original agreement constraint $c(S)\geq \theta$. In particular, setting $\lambda=0$ leads to the original DSP, whereas for $\lambda \to \infty$, the problem prioritizes a node $v$ with the highest agreement value $c_v$. Incrementally increasing $\lambda$ imposes greater penalties on solutions that violate the constraint. Based on the Lagrangian relaxation, we can introduce the Lagrangian dual of \myproblem as follows: 
\begin{align*}
\min_{\lambda\geq 0}\max_{S\subseteq V}\, H_\lambda(S). 
\end{align*}

 The Lagrangian dual aims to find the tightest Lagrangian relaxation of \myproblem: it seeks $\lambda\geq 0$ that minimizes the optimal value of the Lagrangian relaxation (i.e., the upper bound on the optimal value of \myproblem) parameterized by $\lambda$. 

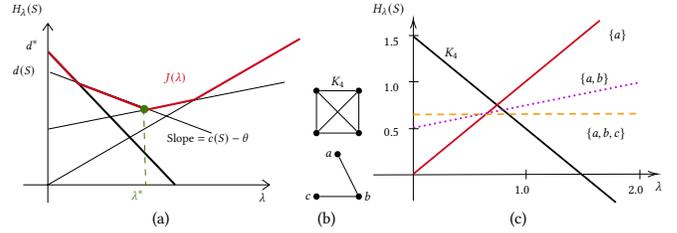
\begin{figure}
\centering
\scalebox{0.70}{
\begin{tabular}{ccc}
\hspace{-6.5mm}
     \resizebox{0.32\textwidth}{!}{%
        \tikzset{every picture/.style={line width=0.75pt}} %set default line width to 0.75pt        

\begin{tikzpicture}[x=0.75pt,y=0.75pt,yscale=-1,xscale=1]

%Shape: Axis 2D [id:dp8402182531292302] 
\draw  (149,205.1) -- (391.8,205.1)(173.28,45.8) -- (173.28,222.8) (384.8,200.1) -- (391.8,205.1) -- (384.8,210.1) (168.28,52.8) -- (173.28,45.8) -- (178.28,52.8)  ;
%Straight Lines [id:da9790261027512657] 
\draw [line width=1.5]    (173.8,74) -- (298.8,205) ;
%Straight Lines [id:da9242158835319259] 
\draw    (175.3,93.9) -- (334.8,154) ;
%Straight Lines [id:da9056714417393594] 
\draw    (421.8,61) -- (173.28,205.1) ;
%Straight Lines [id:da7300470207582994] 
\draw    (173.8,150) -- (406.8,103.8) ;
%Straight Lines [id:da5121141394486566] 
\draw [color={rgb, 255:red, 208; green, 2; blue, 27 }  ,draw opacity=1 ][line width=1.5]    (173.8,74) -- (202.8,105) ;
%Straight Lines [id:da9280130071044845] 
\draw [color={rgb, 255:red, 208; green, 2; blue, 27 }  ,draw opacity=1 ][line width=1.5]    (202.8,105) -- (267.8,129.8) ;
%Straight Lines [id:da507759468227418] 
\draw [color={rgb, 255:red, 208; green, 2; blue, 27 }  ,draw opacity=1 ][line width=1.5]    (270.8,131) -- (317.8,121) ;
%Straight Lines [id:da722098664048229] 
\draw [color={rgb, 255:red, 65; green, 117; blue, 5 }  ,draw opacity=1 ] [dash pattern={on 4.5pt off 4.5pt}]  (267.8,129.8) -- (269.8,205) ;
%Shape: Circle [id:dp06375709615276959] 
\draw  [color={rgb, 255:red, 65; green, 117; blue, 5 }  ,draw opacity=1 ][fill={rgb, 255:red, 65; green, 117; blue, 5 }  ,fill opacity=1 ] (264.8,130.4) .. controls (264.8,128.41) and (266.41,126.8) .. (268.4,126.8) .. controls (270.39,126.8) and (272,128.41) .. (272,130.4) .. controls (272,132.39) and (270.39,134) .. (268.4,134) .. controls (266.41,134) and (264.8,132.39) .. (264.8,130.4) -- cycle ;
%Straight Lines [id:da44415304525158716] 
\draw [color={rgb, 255:red, 208; green, 2; blue, 27 }  ,draw opacity=1 ][line width=1.5]    (317.8,121) -- (421.8,61) ;

% Text Node
\draw (136.56,25) node [anchor=north west][inner sep=0.75pt]    {$H_{\lambda }( S)$};
% Text Node
\draw (380.56,212) node [anchor=north west][inner sep=0.75pt]    {$\lambda $};
% Text Node
\draw (150.56,61) node [anchor=north west][inner sep=0.75pt]    {$d^{*}$};
% Text Node
\draw (289.56,153) node [anchor=north west][inner sep=0.75pt]    {$\mathrm{Slope}=c( S) -\theta $};
% Text Node
\draw (137.56,86) node [anchor=north west][inner sep=0.75pt]    {$d( S)$};
% Text Node
\draw (286.56,93) node [anchor=north west][inner sep=0.75pt]  [color={rgb, 255:red, 208; green, 2; blue, 27 }  ,opacity=1 ]  {$J( \lambda )$};
% Text Node
\draw (254.56,209) node [anchor=north west][inner sep=0.75pt]  [color={rgb, 255:red, 65; green, 117; blue, 5 }  ,opacity=1 ]  {$\lambda ^{*}$};

\end{tikzpicture}
        }
     \hfill\hspace{-5mm} & \resizebox{0.078\textwidth}{!}{%
        \begin{tikzpicture}

  % K_4 graph
  % Nodes in a square
  \node[fill=black, circle, inner sep=3pt] (a) at (0,0) {};
  \node[fill=black, circle, inner sep=3pt] (b) at (2,0) {};
  \node[fill=black, circle, inner sep=3pt] (c) at (2,2) {};
  \node[fill=black, circle, inner sep=3pt] (d) at (0,2) {};

  % Edges
  \draw[thick] (a) -- (b);
  \draw[thick] (a) -- (c);
  \draw[thick] (a) -- (d);
  \draw[thick] (b) -- (c);
  \draw[thick] (b) -- (d);
  \draw[thick] (c) -- (d);

  % Label closer to the graph
  \node at (1,2.4) {\Huge $K_4$};

  % Smaller graph
  % Nodes
  \node[fill=black, circle, inner sep=3pt, label={[font=\Huge]left:{$a$}}] (A) at (1,-1.) {};
  \node[fill=black, circle, inner sep=3pt, label={[font=\Huge]right:{$b$}}] (B) at (2,-3.) {};
  \node[fill=black, circle, inner sep=3pt, label={[font=\Huge]left:{$c$}}] (C) at (0,-3.) {};

  % Edges
  \draw[thick] (A) -- (B);
  \draw[thick] (B) -- (C);

\end{tikzpicture}
        }\hfill\hspace{-5mm}
     & 
     \resizebox{0.32\textwidth}{!}{%
        \tikzset{every picture/.style={line width=0.75pt}} %set default line width to 0.75pt        

\begin{tikzpicture}[x=0.75pt,y=0.75pt,yscale=-1,xscale=1]

%Straight Lines [id:da9699695796085661] 
\draw    (72.5,224.4) -- (390.5,223.41) ;
\draw [shift={(392.5,223.4)}, rotate = 179.82] [color={rgb, 255:red, 0; green, 0; blue, 0 }  ][line width=0.75]    (10.93,-3.29) .. controls (6.95,-1.4) and (3.31,-0.3) .. (0,0) .. controls (3.31,0.3) and (6.95,1.4) .. (10.93,3.29)   ;
%Straight Lines [id:da7260303612545695] 
\draw    (117.5,253.4) -- (117.5,50.4) ;
\draw [shift={(117.5,48.4)}, rotate = 90] [color={rgb, 255:red, 0; green, 0; blue, 0 }  ][line width=0.75]    (10.93,-3.29) .. controls (6.95,-1.4) and (3.31,-0.3) .. (0,0) .. controls (3.31,0.3) and (6.95,1.4) .. (10.93,3.29)   ;
%Straight Lines [id:da9269629206225711] 
\draw [line width=1.5]    (117.5,68.4) -- (345.5,255.4) ;
%Straight Lines [id:da0008478739252026912] 
\draw [color={rgb, 255:red, 208; green, 2; blue, 27 }  ,draw opacity=1 ][line width=1.5]    (117.5,223.4) -- (327.5,50.4) ;
%Straight Lines [id:da09377593838834564] 
\draw [color={rgb, 255:red, 245; green, 166; blue, 35 }  ,draw opacity=1 ][line width=1.5]  [dash pattern={on 5.63pt off 4.5pt}]  (117.5,156.4) -- (370.5,155.4) ;
%Straight Lines [id:da47439325017039136] 
\draw [color={rgb, 255:red, 189; green, 16; blue, 224 }  ,draw opacity=1 ][line width=1.5]  [dash pattern={on 1.69pt off 2.76pt}]  (117.5,171.4) -- (371.5,120.4) ;
%Straight Lines [id:da6723825795685434] 
\draw    (112.5,119.4) -- (123.5,119.4) ;
%Straight Lines [id:da32982618157864096] 
\draw    (112.5,172.4) -- (123.5,172.4) ;
%Straight Lines [id:da4332726970578473] 
\draw    (112.5,67.4) -- (123.5,67.4) ;
%Straight Lines [id:da449520939005434] 
\draw    (244.5,229.4) -- (244.5,218.4) ;
%Straight Lines [id:da8102993513428667] 
\draw    (372.5,229.4) -- (372.5,218.4) ;

% Text Node
\draw (388.56,230) node [anchor=north west][inner sep=0.75pt]  [font=\Large]  {$\lambda $};
% Text Node
\draw (70.56,30) node [anchor=north west][inner sep=0.75pt]  [font=\Large]  {$H_{\lambda }( S)$};
% Text Node
\draw (151.56,80) node [anchor=north west][inner sep=0.75pt]  [font=\Large]  {$K_{4}$};
% Text Node
\draw (335.56,59) node [anchor=north west][inner sep=0.75pt]  [font=\Large]  {$\{a\}$};
% Text Node
\draw (303.56,108) node [anchor=north west][inner sep=0.75pt]  [font=\Large]  {$\{a,b\}$};
% Text Node
\draw (310.56,170) node [anchor=north west][inner sep=0.75pt]  [font=\Large]  {$\{a,b,c\}$};
% Text Node
\draw (83,170) node [anchor=north west][inner sep=0.75pt]  [font=\Large] [align=left] {0.5};
% Text Node
\draw (83,120) node [anchor=north west][inner sep=0.75pt]  [font=\Large] [align=left] {1.0};
% Text Node
\draw (83,68) node [anchor=north west][inner sep=0.75pt]  [font=\Large] [align=left] {1.5};
% Text Node
\draw (230,235) node [anchor=north west][inner sep=0.75pt]  [font=\Large] [align=left] {1.0};
% Text Node
\draw (358,235) node [anchor=north west][inner sep=0.75pt]  [font=\Large] [align=left] {2.0};

\end{tikzpicture}
        }\\
     (a) & (b)& (c) 
\end{tabular}
}
    \caption{(a) Demonstration of \methodlag and (b) a bad instance. 
In the graph in (b), the agreements of nodes $a$, $b$, and $c$ are $1$, $-0.5$, and $-0.5$, respectively. The agreements of nodes in $K_4$ are all $-1$. Let $\theta=0$. The linear relations between $\lambda$ and $H_\lambda(S)$ of some representative subsets are shown in (c). While the optimal solution is $\{a,b,c\}$, the optimal value of $J(\lambda)$ is achieved by only $K_4$ and $\{a\}$. Therefore, \methodlag outputs $\{a\}$, which is a feasible solution but has a density of $0$. 
    }
    \label{fig:lag_demo}
\end{figure}

Let us define $J(\lambda)=\max_{S\subseteq V} H_\lambda(S)$. 
Then we can rewrite the Lagrangian dual as 
\begin{align*}
\min_{\lambda\geq 0}\,J(\lambda).  
\end{align*}
Now observe that for a fixed $S\subseteq V$, $H_\lambda(S)$ is a linear function of $\lambda$. Specifically, the line intersects the y-axis at $d(S)$ with a slope of $c(S) - \theta$, as shown in Figure~\ref{fig:lag_demo}(a). 
Hence, we see that 
$J(\lambda)$ is a piece-wise linear, convex function. 
Let $\Lambda^*$ be the set of optimal solutions to the Lagrangian dual, 
i.e., $\Lambda^*= \argmin_{\lambda\geq 0}J(\lambda)$. 
Then it is easy to see that for any $\lambda < \min \Lambda^*$, any optimal solution to the Lagrangian relaxation with $\lambda$ is not feasible for \myproblem, whereas for any $\lambda > \min \Lambda^*$, the above is feasible for \myproblem. 
Furthermore, for $\lambda=\min\Lambda^*$, there is at least one optimal solution $S^*\subseteq V$ to the Lagrangian relaxation with $\lambda$ that is feasible for \myproblem. 
From the above, $S^*$ (or equivalently, any optimum to the Lagrangian relaxation with any other $\lambda \in \Lambda^*$ if exists) can be seen as a solution of \myproblem that respects the density $d(S)$ as much as possible (because of the minimality of $\lambda$) while satisfying the agreement constraint $c(S)\geq\theta$, among all optimal solutions to the Lagrangian relaxations with all $\lambda\geq 0$.

Therefore, the goal of our algorithm is to solve the Lagrangian dual, i.e., to compute an element $\lambda^*\in \Lambda^*$ 
and the corresponding optimal solution $S^*\subseteq V$ to the Lagrangian relaxation with $\lambda^*$. 
This can be done using binary search over $\lambda\geq 0$ with an arbitrarily small additive error $\epsilon>0$, based on the aforementioned piece-wise linearity and convexity of $J(\lambda)$. Indeed, we can conduct binary search over $\lambda\geq 0$ with respect to the sign of the slope of $J(\lambda)$ at the point. 

\begin{algorithm}[t]
\caption{ \methodlag }\label{alg:lag_flow}
\SetKwInOut{Input}{Input}
\SetKwInOut{Output}{Output}
\Input{\ $G=(V,E)$, $\bp_v\in \bbR^d$ for each $v\in V$, $\bq\in \bbR^d$, $\theta \in \bbR$, and precision parameter $\epsilon>0$}
\Output{\ $S\subseteq V$}
$d^*\leftarrow \max_{S\subseteq V}\frac{|E(S)|}{|S|}$, $c_{\max} \leftarrow  \max_{v\in V}c_v$\;
$\lambda^L\leftarrow 0$, $\lambda^R\leftarrow \frac{d^*}{c_{\max}-\theta}$\;
$S_\text{out}\leftarrow \text{opt. sol. to the Lagrangian relaxation with } \lambda^R$\;
\While{$\lambda^R-\lambda^L >\epsilon$}{
$\lambda\leftarrow (\lambda^R+\lambda^L)/2$\;
Let $S$ be an optimal solution to the Lagrangian relaxation with $\lambda$ (i.e., HDSP-PN with $(G,\bm{w})$, where $\bm{w}=(w_v)_{v\in V}$ such that $w_v=\lambda\cdot c_v$ for $v\in V$)\;
\textbf{if} $c(S)\geq \theta$ \textbf{then} $S_\text{out}\leftarrow S$, $\lambda^R\leftarrow \lambda$\;
\textbf{else} $\lambda^L\leftarrow \lambda$\;
}
\Return $S_\text{out}$\;
\end{algorithm}

The pseudocode is described in Algorithm~\ref{alg:lag_flow}. The search range of $\lambda$ is upper bounded by $\frac{d^*}{c_{\max}-\theta}$, as the optimal value of the Lagrangian relaxation with $\lambda=\frac{d^*}{c_{\max}-\theta}$ is at least that with $\lambda=0$, i.e., $d^*=\max_{S\subseteq V}\frac{|E(S)|}{|S|}$, by taking nodes with the largest agreement. 
Letting $t_{\text{HDSP-PN}}(G,\bm{w})$ be the time complexity required to solve HDSP-PN with $(G,\bm{w})$, which is known to be strongly polynomial in $n$ and $m$ (i.e., $O(n)$ max-flow computations)~\cite{huang2023densest}, the complexity of Algorithm~\ref{alg:lag_flow} can be evaluated as 
$O\left(t_{\text{HDSP-PN}}(G,\bm{w})\log\frac{d^*}{\epsilon(c_\text{max}-\theta)}\right)$. 

As stated in Proposition~\ref{prop:eth_hardness}, \myproblem is hard to approximate. Consistent with this, the output of Algorithm~\ref{alg:lag_flow} can be arbitrarily bad in terms of the multiplicative factor. An extreme scenario is where the function $J(\lambda)$ is overshadowed by an infeasible densest subgraph and an isolated node with the high agreement value, leading to only the isolated node being returned as a feasible solution. An illustrative example of such a graph is shown in Figure~\ref{fig:lag_demo}(b), with the corresponding $H_\lambda(S)$ functions depicted in Figure~\ref{fig:lag_demo}(c). While the optimal solution is the subset $\{a, b, c\}$ with density $2/3$, the result from Algorithm~\ref{alg:lag_flow} would be $\{a\}$ with density $0$. 
Nevertheless, we can prove an a-posteriori, additive approximation guarantee as follows. 
The proof can be found in Appendix~\ref{apdx:prop4}.

\begin{proposition}\label{prop:lagr}
Algorithm~\ref{alg:lag_flow} (\methodlag) outputs 
a feasible solution $S_\mathrm{out}\subseteq V$ for \myproblem that satisfies 
\begin{align*}
d(S_\mathrm{out})\geq \mathrm{OPT} - \lambda^R(c(S_\mathrm{out}) - \theta), 
\end{align*}
where $\mathrm{OPT}$ is the optimal value of \myproblem and $\lambda^R$ is its value at the termination of the algorithm. 
\end{proposition}
The above proposition implies that the closer the agreement value $c(S_\text{out})$ is to $\theta$, the more near-optimal the output $S_\mathrm{out}$ is. This measure serves as a useful indicator for practitioners, assisting them in determining whether the obtained result is sufficiently reliable or there is a need to pursue further analysis.

\subsection{LP dual-inspired greedy peeling algorithm}\label{subsec:dual}

Our algorithm (\method) adopts a greedy peeling algorithm, inspired by the dual of an LP relaxation. 
The algorithm first looks at the entire node set $V$ and iteratively removes a node with the lowest priority. 
Assume that the current node subset kept by the algorithm is $V'\subseteq V$. Intuitively, we want to remove a node that is sparsely connected to the others in $V'$ and associated with a small agreement value. To facilitate this, we introduce the concept of node's \textit{load}, defined based on both its degree in $G[V']$ and its agreement value. Formally, the load of node $v\in V'$ is defined as 
\begin{align}\label{eq:load}
\ell_v=\degree_{V'}(v) + z_2(c_v-\theta),
\end{align}
where $\deg_{V'}(v)$ is the degree of $v\in V'$ on the induced subgraph $G[V']$ and $z_2$ is a nonnegative parameter that enables to control the trade-off between the degree and the agreement value, which will be justified later using the dual of an LP relaxation. 
For convenience, we   define the load of $V'$ as 
\begin{equation}\label{eq:load_subset}
\ell(V')=\min_{v\in V'}\ell_v. 
\end{equation}
For a fixed value of $z_2$, our algorithm iteratively removes a node with the least load until only one node is left. 
The best feasible solution identified during this peeling procedure, which maximizes the density while satisfying the agreement constraint, is then selected (if it exists) as the candidate for the fixed $z_2$. 

\begin{algorithm}[t]
\caption{ \method}
\label{alg:peel}
\SetKwInOut{Input}{Input}
\SetKwInOut{Output}{Output}
\Input{\ $G=(V,E)$, $\bp_v\in \bbR^d$ for each $v\in V$, $\bq\in \bbR^d$, $\theta \in \bbR$, and precision parameter $\epsilon>0$}
\Output{\ $S\subseteq V$}
$\degree_{\max} \leftarrow \max_{v\in V}\degree_V(v)$\; $\delta_{\min}\leftarrow \min_{u,v\in V:\,c_u\neq c_v}|c_u-c_v|$\;
$z_2^L\leftarrow 0$, $z_2^R\leftarrow 2\cdot \degree_{\max}/\delta_{\min}$\;
$T\leftarrow \argmax_{v\in V}c_v$\;
$S_\mathrm{out}\leftarrow \argmax_{v\in V}c_v$ \\ \qquad \quad \ (i.e., a solution corresponding to $T$ of $z_2^R$)\;
\While{$z_2^R-z_2^L>\epsilon$}{
  $z_2\leftarrow (z_2^L+z_2^R)/2$\;
  $V_n \leftarrow V$, $i\leftarrow n$, $S_\mathrm{cand}\leftarrow \emptyset$\;
  \While{$i>1$}{
    Take $v_\text{min} \in \argmin_{v\in V_i} (\degree_{V_i}(v)+z_2(c_v-\theta))$\;
    $V_{i-1}\leftarrow V_i\setminus \{v_\text{min}\}$, $i\leftarrow i-1$\;
  }
  $T\leftarrow \text{any element of } \argmax\{\ell(S)\mid S\in \{V_n,\dots, V_1\}\}$\;
  \If{$c(T)\geq \theta$}{
    $z_2^R\leftarrow z_2$\;
    $S_\mathrm{cand}\leftarrow \argmax\{d(S)\mid S\in \{V_n,\dots, V_1\},\, c(S)\geq \theta\}$\;
    \textbf{if} $d(S_\mathrm{cand})> d(S_\mathrm{out})$ \textbf{then} $S_\mathrm{out}\leftarrow S_\mathrm{cand}$\;
  }
  \textbf{else} $z_2^L\leftarrow z_2$\;
}
\Return $S_\mathrm{out}$\;
\end{algorithm}

Determining an appropriate value of $z_2$ is a critical component of our algorithm. 
To this end, we employ a binary search, which is guided by the feasibility of a specific node subset $T\subseteq V$ that achieves the highest load during the peeling procedure. If $T$ is feasible, i.e., $c(T)\geq \theta$ holds, then we decrease the value of $z_2$; otherwise, we increase the value of $z_2$ to find a feasible $T$. The pseudocode of our algorithm is detailed in Algorithm~\ref{alg:peel}. In the algorithm, the upper bound on the search range of $z_2$ is given by $2\cdot \deg_{\max}/\delta_{\min}$. This is because of the fact that any $z_2$ greater than or equal to this value makes the peeling procedure remove the nodes in the increasing order of their agreement values. 
As the single iteration of the while-loop just takes $O(m+n\log n)$ time, like the standard greedy peeling algorithm for DSP on edge-weighted graphs, and the number of iterations is upper bounded by $O\left(\log\left(\frac{\deg_{\max}/\delta_{\min}}{\epsilon}\right)\right)$, Algorithm~\ref{alg:peel} runs in $O\left((m+n\log n)\log\left(\frac{\deg_{\max}/\delta_{\min}}{\epsilon}\right)\right)$ time.

It should be remarked that the feasibility of $T$ is not necessarily \emph{monotonic} with respect to $z_2$. 
Specifically, it is not necessarily guaranteed that there exists some value of $z_2$ such that $T$ is always feasible (resp. infeasible) for larger (resp. smaller) $z_2$. 
Therefore, our binary search can be seen as a heuristic method that aims to find the smallest $z_2$ for which $T$ is feasible. 
However, taking into account the role of $z_2$, it can be expected that the monotonicity of the feasibility of $T$ usually holds in practice. 

From now on, we interpret Algorithm~\ref{alg:peel} through the lens of the dual of an LP relaxation, based on which we present the theoretical guarantee of the algorithm. Following the LP relaxation of DSP by Charikar~\cite{Charikar00}, we can introduce an LP relaxation of \myproblem: 
\begin{equation}\label{primal}
    \begin{array}{ll@{}ll}
    \text{max.} & \sum_{e\in E} x_e  \\
    \text{s.t.} & \sum_{v\in V} y_v\leq 1, \\
    & \sum_{v\in V} c_v\cdot y_v \geq \theta, \\
    & x_e\leq y_u,\ \ x_e\leq y_v& & \forall e=\{u,v\}\in E. \\
    \end{array}
\end{equation}
Note that the LP~\eqref{primal} has an additional constraint corresponding to the agreement constraint, compared with the LP for DSP. 
The LP~\eqref{primal} is indeed a relaxation of \myproblem: 
Take an optimal solution $S^*\subseteq V$ to \myproblem. If we construct a solution of the LP~\eqref{primal} as $x_e=\frac{1}{|S^*|}$ for each $e\subseteq S^*$ and 0 for the others; $y_v=\frac{1}{|S^*|}$ for each $v\in S^*$ and 0 for the others, then the solution is clearly feasible. Moreover, the objective value then equals $\frac{|E(S^*)|}{|S^*|}=d(S^*)$, i.e., the optimal value of \myproblem.
The dual of the LP can be written as 
\begin{equation}\label{dual}
    \begin{array}{ll@{}ll}
    \text{min.} & z_1 - \theta \cdot z_2  \\
    \text{s.t.} & z_1 -\theta\cdot z_2   \geq (c_v-\theta)z_2 + \sum_{e\in E:v\in e} f_v^e  & & \forall v\in V, \\
    & f_u^e + f_v^e \geq 1 & & \forall e=\{u,v\}\in E, \\
    & f_u^e, f_v^e \geq 0 & & \forall e=\{u,v\}\in E, \\
    & z_1,z_2\geq 0.
    \end{array}
\end{equation}
Note that for the sake of presentation, the term $\theta\cdot z_2$ is subtracted from both sides of the first constraint, which does not essentially change the problem. By the strong duality of LP, the optimal value to the dual LP~\eqref{dual} is equal to the optimal value of the primal LP~\eqref{primal}, i.e., an upper bound on the optimal value of \myproblem. 

Similarly in the case of DSP~\cite{Charikar00}, the dual LP~\eqref{dual} with a fixed value of $z_2$ can be seen as a \emph{load balancing} problem, 
where we are asked to allocate the unit load of each edge $e=\{u,v\}\in E$ to its endpoints $u,v$ (fractionally) so as to minimize the maximum load allocated (in addition to the quantity led by the agreement value, i.e., $(c_v-\theta)z_2$) over nodes. 
Each iteration of Algorithm~\ref{alg:peel},  corresponding to a concrete value of $z_2$, can be seen as an algorithm to compute a feasible solution for the dual LP~\eqref{dual} (with the fixed $z_2$). 
Indeed, when removing a node $v_{\min}$, we can allocate all unit loads of its incident edges to $v_{\min}$. 
Then we see that the right-hand-side of the first constraint in the dual LP~\eqref{dual} for that solution is equal to the load $\ell_v$, defined in \eqref{eq:load}, calculated when $v$ is removed, for any $v\in V$. 
By setting $z_1$ so that it satisfies the first constraint, we obtain a feasible solution for the dual LP~\eqref{dual} whose objective value equals $\ell(T)$, defined in~\eqref{eq:load_subset}. 
Therefore, we see that $\ell(T)$ is an upper bound on the optimal value of \myproblem. 

Recall that for DSP, the dual LP~\eqref{dual} and the definitions of loads in~\eqref{eq:load} and~\eqref{eq:load_subset} reduce to those with $z_2=0$, respectively~\cite{Charikar00}. Therefore, the load $\ell(T)$ is concerned only with the degrees of $v\in T$, and it is easy to see that $d(T)\geq \ell(T)/2$ holds, implying that $T$ is a $1/2$-approximate solution. 
However, for \myproblem, the load $\ell(T)$ also involves the agreement values of $v\in T$, which makes it difficult to directly obtain such an approximation.

We can again prove an a-posteriori approximation guarantee. 
The proof can be found in Appendix~\ref{apdx:prop5}. 

\begin{proposition}\label{prop:theory_dual}
Algorithm~\ref{alg:peel} (\method) outputs a feasible solution $S_\mathrm{out}\subseteq V$ for \myproblem that satisfies 
\begin{align*}
d(S_\mathrm{out})\geq \frac{\mathrm{OPT}}{2} - \frac{z_2^R}{2}(c(T)-\theta), 
\end{align*}
where $\mathrm{OPT}$ is the optimal value of \myproblem, $z_2^R$ is its value at the termination of the algorithm, and $T$ is the last subset $T$ found by the algorithm that satisfies $c(T)\geq \theta$. 
\end{proposition}

\section{Experiments}
\label{sec:exp}

 In this section, we evaluate the performance of our proposed algorithms, \methodlag and \method, in terms of both solution quality and running time. 
 We also demonstrate the practical usefulness of our methodology through a variety of case studies.

\subsection{Experimental setup}

\noindent \textbf{Datasets.} 
We employ the following three graphs associated with real feature information whose statistics are summarized in Table~\ref{tab:datasum}. 
\begin{itemize}
 \leftskip=-15pt
 
  \item \textit{Twitter}. We have garnered a  dataset from Twitter that includes a network of users {\it follow} interactions as well as their views on two timely topics: vaccination against COVID-19  and the Ukraine war. These opinions are evaluated and given scores ranging from $-1$ and $+1$ by GPT-3.5, outlined in Appendix~\ref{appendix:twitter}. Opinion close to $+1$ means that the user has positive opinion on vaccination and supports Ukraine, whereas opinion close to $-1$ implies that the user is against vaccination and does not support Ukraine.

    \item \textit{DBLP}~\cite{Miyauchi+23}. 
    We use a network from the authors of papers presented at major conferences across six areas (Theory, Data Mining, Image Processing, Learning, Data Management, and Networking) from 2003 to 2022. In this graph, nodes represent authors with their attributes indicating  the areas in which they have published. Edges between nodes show that the corresponding authors have collaborated in at least one occasion. In order to assess researcher's expertise in a particular area, we calculate $\lfloor \log(1+k) \rfloor$, where $k$ represents the total number of publications the researcher has in that area.  After this processing, we obtain a network where each node has a six-dimensional vector with those expertise values.      

    \item \textit{Deezer-HR}~\cite{rozemberczki2019gemsec}. Nodes represent users on a music streaming platform Deezer and edges represent mutual friendships. For each user, the dataset also provides its preference on 84 music types and we formulate it as a binary vector, with an entry being $+1$ if the corresponding music type is liked and $0$ otherwise. 
 
    \end{itemize}

\begin{table}[t]
\centering
	\caption{Statistics of datasets for case study.}
	\label{tab:datasum}
    \scalebox{0.75}{
	\begin{tabular}{c|c|c|c|c}\toprule
		Name & $n$ & $m$   & Feature type & Dimension \\
		\midrule
		Twitter & 3\,409 & 11\,508 & Political opinion & 2 \\
            DBLP & 25\,714 & 152\,100 & Research expertise& 6 \\
            Deezer-HR & 54\,573 & 498\,202 & Music preference & 84 \\
\midrule
                Amazon & 334\,863 & 825\,872 & Synthetic $c_v$'s& ---\\ 
    DBLP-SNAP & 317\,080 & 1\,049\,866 & Synthetic $c_v$'s & --- \\
    YouTube & 1\,134\,890 & 2\,987\,624 & Synthetic $c_v$'s & --- \\
    LiveJournal & 3\,997\,962 & 34\,481\,189 & Synthetic $c_v$'s & --- \\
            \bottomrule
	\end{tabular}
    }
\end{table}

In addition, we use four larger real-world graphs collected from SNAP~\cite{snap}, up to a graph with more than 34 million edges, to further examine the scalability of the algorithms. 
Table~\ref{tab:datasum} also summarizes the basic statistics of the graphs. 
While the graphs do not initially come with opinions, we randomly sample the agreement values for half of the nodes from $\mathcal{N}(0.1,0.01)$ and for the other half of nodes from $\mathcal{N}(-0.1,0.1)$.

\smallskip
\noindent \textbf{Baselines.} 
We use the following three heuristics as our baselines:

\begin{itemize}
\leftskip=-15pt
\item \textit{Agreement Filtering (AF)}. This method first removes all  nodes whose agreement value is less than the threshold $\theta$, and then returns the densest subgraph in the remaining graph. 
\item \textit{LP and Sweep (LP-S)}. We first  solve the LP~\eqref{primal}  to get an optimal solution $((x^*_e)_{e\in E},(y^*_v)_{v\in V})$. Then for each unique value $y\in \{y^*_v\}_{v\in V}$, we create a node subset $S_{y}=\{v\mid y^*_v\geq y\}$ and check if it satisfies the constraint. The feasible subset with the largest density is returned if it exists. The single node with the largest agreement is returned if no feasible solution is found. 
\item \textit{LP and Greedy (LP-G)}. This baseline also solves the LP~\eqref{primal} first. Following this, it creates a subgraph consisting of all nodes that have positive values for the variables in the LP solution, i.e., $S=\{v\mid y^*_v>0\}$. We next iteratively remove the node with the smallest agreement value, until the subset at hand satisfies the constraint. We return the resulting subset if it is not empty; otherwise, we return the node with the largest agreement value.  
\end{itemize}

While our problem can be formulated through integer linear programming, it is impractical to solve. 
For instance, it cannot be solved in two hours on a 100-node graph.

\smallskip 
\noindent \textbf{Machine specs and code.} The experiments are performed on a single machine, with Intel i7-10850H CPU @ 2.70GHz and 32GB of main memory. The main algorithms are implemented in C++. LPs are coded and solved via Gurobi Optimizer's Python API.
The code is publicly available at https://github.com/tsourakakis-lab/q-disco.

\begin{figure*}[!ht]
    \centering
    \begin{subfigure}[]{.17\linewidth}
        \centering
        \includegraphics[width=\linewidth]{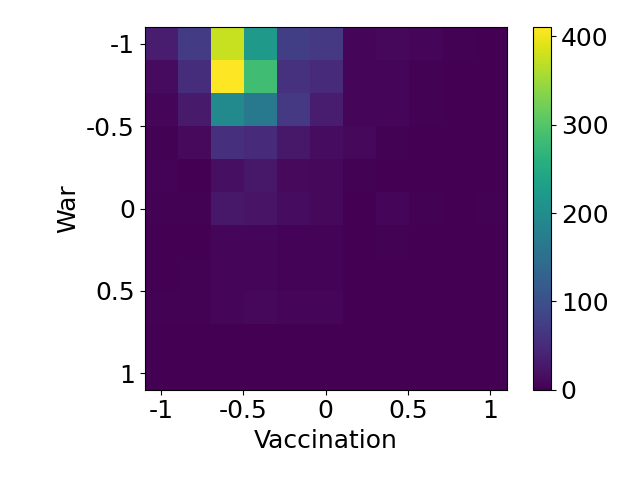}
        \vspace{-6mm}
        \caption{}
    \end{subfigure} \ \ 
    \begin{subfigure}[]{.17\linewidth}
        \centering
        \includegraphics[width=\linewidth]{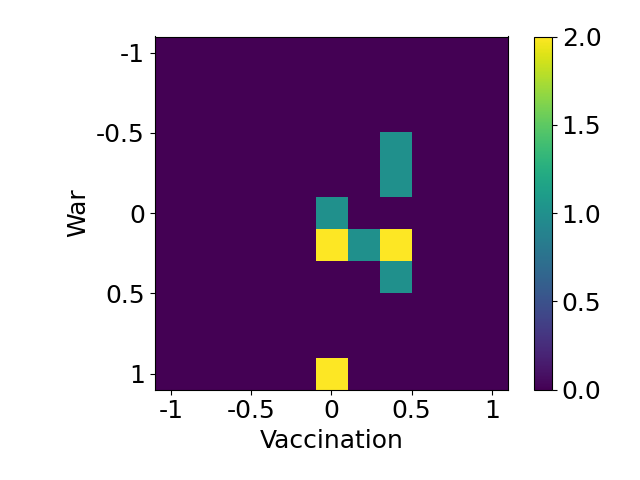}
        \vspace{-6mm}
        \caption{}
    \end{subfigure}\ \ 
    \begin{subfigure}[]{.17\linewidth}
        \centering
        \includegraphics[width=\linewidth]{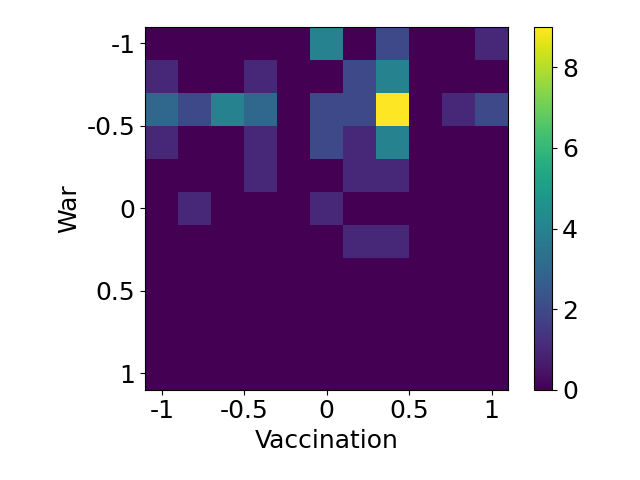}
        \vspace{-6mm}
        \caption{}
    \end{subfigure}\ \ 
    \begin{subfigure}[]{.17\linewidth}
        \centering
        \includegraphics[width=\linewidth]{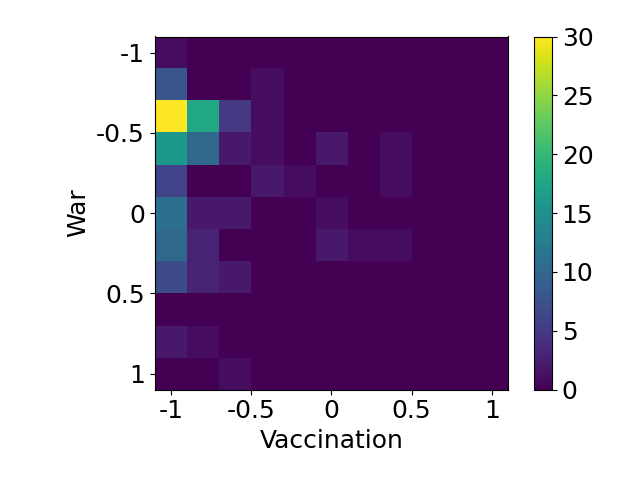}
        \vspace{-6mm}
        \caption{}
    \end{subfigure}\ \ 
    \begin{subfigure}[]{.17\linewidth}
        \centering
        \includegraphics[width=\linewidth]{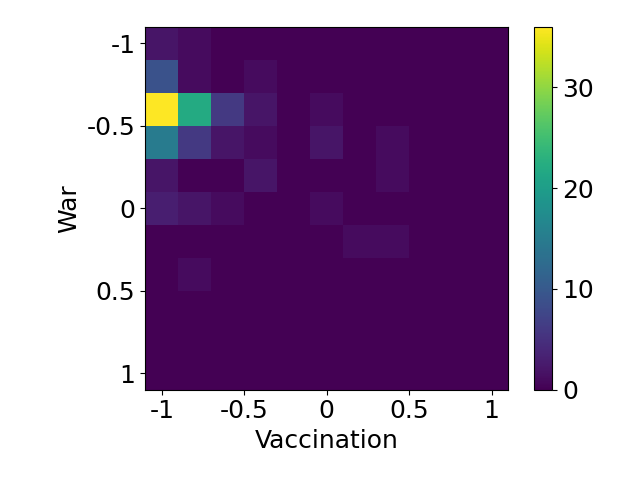}
        \vspace{-6mm}
        \caption{}
    \end{subfigure} 
    \caption{\label{fig:twitter} Histogram of opinions on $(\textsf{Vax}, \textsf{Ukraine})$ for the (a) full dataset and the outputs of \method for $\theta=0.5$ with queries (b) $\bq=(1,1)$, (c) $\bq=(1,-1)$, (d) $\bq=(-1,1)$, and (e) $\bq=(-1,-1)$.
    }
\end{figure*}

\subsection{Results for Twitter dataset}
\label{sec:twitter}

\begin{table}[]
    \centering
    \caption{  \label{tab:performancetwitter} Performance of algorithms on the Twitter dataset.}
    \scalebox{0.69}{ 
    \begin{tabular}{c|c|c|c|c|c|c|c|c|c|c|c}\toprule
    & \multicolumn{2}{c|}{AF }  & \multicolumn{2}{c|}{LP-S } & \multicolumn{2}{c|}{LP-G } & \multicolumn{2}{c|}{\methodlag } & \multicolumn{2}{c|}{\method } & \\
    Query  &  $d(S)$ &  $c(S)$  & $d(S)$ &  $c(S)$ &  $d(S)$ &  $c(S)$ &  $d(S)$ &  $c(S)$ & $d(S)$ &  $c(S)$ & UB\\ \midrule
    $(1,1)$ & 0 & 0.72 & 0 &	 1.20 &	 1.08 &	 0.54 &  0 & 1.20 & 2.29 & 0.51  & 7.7\\
    $(1,-1)$ & 0.83 & 0.80 & 0 &	 2.00 &	 2.23 &	 0.52 & 0 & 2.00 & 7.58 & 0.50  & 11.7\\ 
    $(-1,1)$ & 5.03 & 0.73 & 0 &	 1.80 &	 10.66 &	 0.50 & 0 & 1.80  & 14.83 & 0.50 & 19.5 \\ 
    $(-1,-1)$ & 11.55 & 1.25 & 18.68 &	 0.69 &	 18.68 &	 0.69 & 18.67 & 0.71 & 18.20 & 0.86 & 18.88 \\ \bottomrule
    \end{tabular}
    }
\end{table}

\noindent
\textbf{Performance test.}
Let $\bq=(\textsf{Vax},\textsf{Ukraine})$ be a query vector, where \textsf{vax} is the opinion for vaccination while \textsf{Ukraine} is the opinion for the Ukraine war. Setting each of the four possible polarity vectors $\bq=(1,1),(1,-1),(-1,1),(-1,-1)$ as a query, we run the algorithms with the threshold $\theta=0.5$. 

The results are shown in Table~\ref{tab:performancetwitter}, 
where we also report the upper bound on the optimal value obtained by Proposition~\ref{prop:lagr}, denoted by UB.
As can be seen, \method outperforms the baseline methods and even \methodlag. 
Indeed, for the first three queries $\bq=(1,1),(1,-1),(-1,1)$, \method achieves much larger densities than those of the outputs of the other methods, and for the last query $\bq=(-1,-1)$, all methods except AF find near-optimal solutions according to the upper bound. 
As for the efficiency, both LP baselines take around five seconds to finish, while \method and \methodlag terminate in 0.02 and 0.14 seconds, respectively.
It should be noted that AF performs poorly over all queries. 
This is expected, as the method is too conservative: the output cannot contain any node with an agreement value lower than the threshold $\theta$. 
Moreover, LP-S and \methodlag often output a single node with the largest agreement, resulting in $d(S)=0$, as they fail to find a (meaningful) feasible solution. 
This is due to the fact that 
as will be shown later, the opinions of nodes are biased towards negative opinions, at least partially contradicting the first three queries. 
In Appendix~\ref{appendix:experiments_twitter}, we also present the results of our performance test when we vary the value of $\theta$. 

\smallskip
\noindent
\textbf{Case study.} Figure~\ref{fig:twitter} 
and Table~\ref{tab:resultstwitter} 
offer a comprehensive analysis of our findings from the Twitter dataset; in particular, Figure~\ref{fig:twitter} showcases feature empirical histograms detailing the distribution of opinions regarding the topics of vaccination and the Ukraine war. Specifically, Figure~\ref{fig:twitter}(a) shows the opinions across all users in our dataset. It is worth outlining that the crawling was performed starting from a self-declared Republican with strong negative opinions on both topics, thus creating a dataset with these characteristics. 
Figures~\ref{fig:twitter}(b)--(e) show the corresponding 
histograms for the outputs of \method obtained in the performance test. The histograms confirm that the opinions of most nodes in the outputs are consistent with the given queries, except for the case of $\bq=(-1,1)$ (Figure~\ref{fig:twitter}(d)), resulting in an average negative opinion even on the Ukraine war. To achieve a positive opinion on the war, we can adjust the $\theta$ value upwards, by utilizing the flexibility of our methodology. 
For instance, if we set $\theta=1.1$, we can obtain $S\subseteq V$ with $|S|=50$ and $|E(S)|=141$ with average \textsf{Vax} of $-.91\pm.15$ and average \textsf{Ukraine} of $.21\pm.40$, aligning with the query. The histogram of opinions of this subgraph can be found in Appendix~\ref{appendix:experiments_twitter}. 
Finally, we also highlight that for $q=(1,1)$, the resulting subgraph 
does have positive average opinions on both topics (see Table~\ref{tab:resultstwitter}), despite the fact that the whole dataset is biased towards negative opinions. It can be inferred that we have already found nearly the most positive but tightly connected group in this graph. One supporting evidence is that the baseline AF returns an empty set with $\theta=0.5$, indicating no highly positive nodes connecting each other in the dataset.

\begin{table}[]
    \centering
    \caption{  \label{tab:resultstwitter} 
    Details of the outputs of \method with $\theta=0.5$.
    }
    \scalebox{0.7}{ 
    \begin{tabular}{c|c|c|cc|cc}\toprule
    Query  & $|S|$ & $|E(S)|$ & \multicolumn{2}{c|}{Vacc. op} & \multicolumn{2}{c}{Ukr. op} \\ 
    $\bq=(\textsf{Vax}, \textsf{Ukr})$ & & & avg & std & avg & std \\ \midrule
    Global & 3.4K & 11.5K & -.51 & .39 & -.30 & .27 \\ 
    $ (1,1)$ & 17 & 39 & .38 & .30 & .37 & .38 \\ 
    $ (1,-1)$ & 88 & 667 & .16 & .56 & -.52 &  .29 \\ 
    $ (-1,1)$ & 346 & 3364 & -.84 & .25 & -.12 & .34 \\
    $ (-1,-1)$ & 145 & 2472 & -.76 & .34 & -.42 & .24 \\ \bottomrule
    \end{tabular}
    }
\end{table}

\begin{figure*}[!ht]
    \centering
    \begin{subfigure}[]{.19\linewidth}
        \centering
        \includegraphics[width=\linewidth]{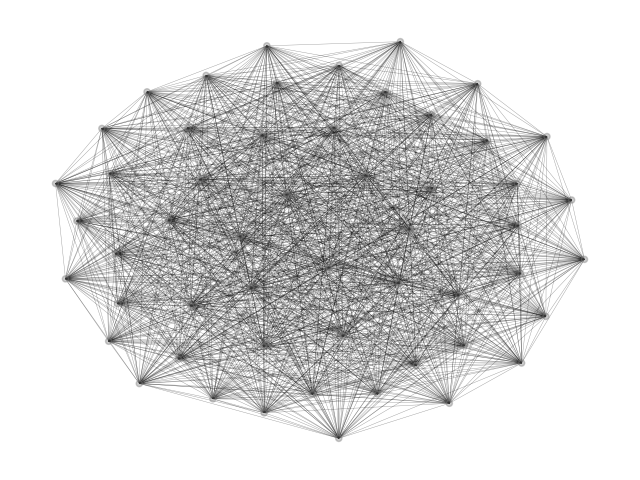}
        \vspace{-6mm}
        \caption{}
    \end{subfigure} \quad
    \begin{subfigure}[]{.19\linewidth}
        \centering
        \includegraphics[width=\linewidth]{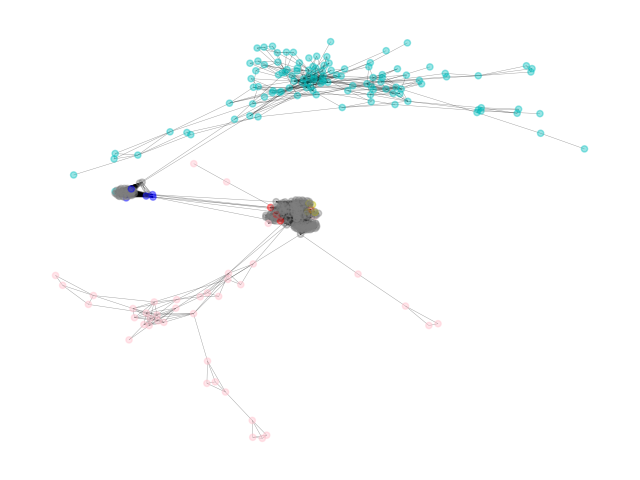}
        \vspace{-6mm}
        \caption{}
    \end{subfigure}\quad
    \begin{subfigure}[]{.19\linewidth}
        \centering
        \includegraphics[width=\linewidth]{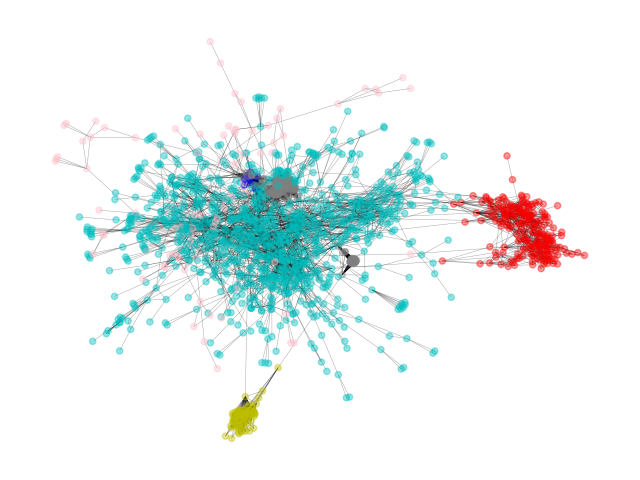}
        \vspace{-6mm}
        \caption{}
    \end{subfigure}\quad
    \begin{subfigure}[]{.19\linewidth}
        \centering
        \includegraphics[width=\linewidth]{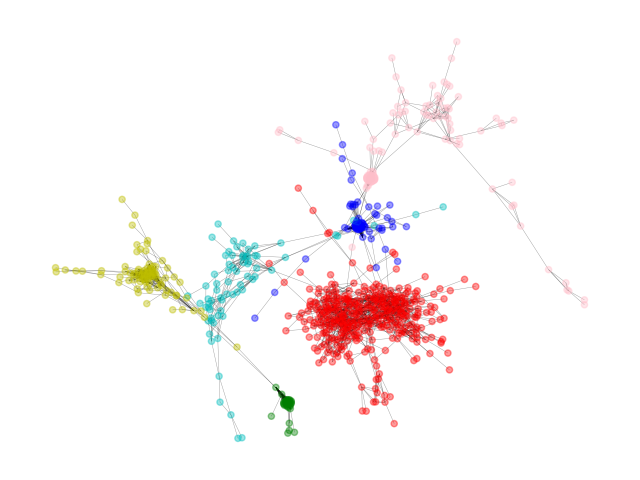}
        \vspace{-6mm}
        \caption{}
    \end{subfigure}\quad
    \begin{subfigure}[]{.12\linewidth}
        \centering
        \includegraphics[width=\linewidth]{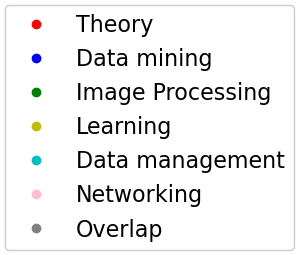}
        \caption{}
    \end{subfigure} 
    \caption{\label{fig:dblp}
 Results for the DBLP dataset with adjusting the threshold $\theta$:  union of the outputs for queries $q_a$ over all areas, with the thresholds (a) $\theta=-100$, (b) $\theta=-1$, (c) $\theta=0$, and (d) $\theta=1.5$. The coloring of nodes is based on the specific query that extracts them, as detailed in (e). Notably, a node is colored gray if it is found by more than one query.}
\end{figure*}

\subsection{Results for DBLP dataset}

\noindent
\textbf{Performance test.} 
For each area $a$ of the six areas encompassed by the DBLP dataset, we define a query vector as 
\begin{align*}
\bq_{a} = \bm{e}_a - \sum_{\text{all other areas~}a' \neq a} \bm{e}_{a'},
\end{align*}
where $\bm{e}_a$ denotes the unit basis vector that has $1$ for the corresponding area $a$. Intuitively, each query encapsulates a focused differentiation strategy in the multi-dimensional space, where each dimension corresponds to a different area of interest. The overall effect of this vector formulation is to create a contrast between the area of interest and all other areas. 
We perform the algorithms with the threshold $\theta=0$. 

The results are presented in Table~\ref{tab:resultsDBLP}. 
The trend is the same as that for the Twitter dataset: 
\method outperforms the baseline methods and \methodlag, and obtains near-optimal solutions for most queries. 
The superiority of \method and \methodlag in terms of efficiency becomes drastic for this larger dataset. 
Indeed, \method and \methodlag run in 0.34 and 2.90 seconds, respectively, on average, while the LP baselines take more than 242 seconds.

\begin{table}[]
    \centering
    \caption{Performance of algorithms on the DBLP dataset.}
    \label{tab:resultsDBLP}
    \scalebox{0.62}{
    \begin{tabular}{c|c|c|c|c|c|c|c|c|c|c|c}
    \toprule
    & \multicolumn{2}{c|}{AF }  & \multicolumn{2}{c|}{LP-S }& \multicolumn{2}{c|}{LP-G }  & \multicolumn{2}{c|}{\methodlag } & \multicolumn{2}{c|}{\method } & \\
        Query &  $d(S)$ &  $c(S)$ &  $d(S)$ &  $c(S)$ &  $d(S)$ &  $c(S)$ &  $d(S)$ &  $c(S)$&  $d(S)$ &  $c(S)$ & UB \\\midrule
        Theory &  4.73 & 1.71 &1.86 & 3.00 &  3.41 & 0.04 &  1.86 & 3.00 & 13.28 & 0.00 & 14.28 \\
        Data Mining &  6.00 & 1.15 &21.00 & 0.02 &  10.93 & 0.02&  21.0 & 0.02 & 20.00 & 0.22 & 21.05\\
        Image Processing &  10.29 & 1.17 &25.91 & 0.01 &  25.89 & 0.01&  25.91 & 0.01 & 25.09 & 0.01 & 25.97\\
        Learning &  10.92 & 1.49 & 19.78 & 0.26 &  19.78 & 0.26&  9.23 & 1.93 & 20.83 & 0.01& 21.56 \\
        Data Management &  3.69 & 1.31 & 0.50 & 3.00 &  0.62 & 0.12&  0.50 & 3.00 & 8.97 & 0.00 & 13.88 \\
        Networking &  5.56 & 1.49 & 0.50 & 4.00 &  2.6 & 0.00 &  0.50 & 4.00 & 15.00 & 0.00 & 16.46 \\
    \bottomrule
    \end{tabular}
    }
    \label{tab:dblp_theta_neg1}
\end{table}

\smallskip
\noindent
\textbf{Case study.}
We demonstrate that by adjusting $\theta$ in a wide range, we can see transitioning from the densest subgraph, where the constraint is effectively negligible because it is trivially met, to a subgraph that solely comprises experts from the targeted area.
Within the intermediate range of $\theta$, we aim to reveal the intersecting architecture of dense subgraphs, where it is possible that some nodes contributing to the density might not align with the query vector. We verify our findings in Figure~\ref{fig:dblp}. The figure illustrates the combination of subgraphs obtained by \method for each query vector used in the performance test. It is observed that at a $\theta$ value as low as $-100$, the result is essentially the densest subgraph. When $\theta$ falls within the range of $[-1, 1]$, distinct communities emerge with significant overlap among them. As $\theta$ increases to 1.5, the communities become nearly disjoint, representing authors who focus primarily on their specific field.

\begin{table*}[]
    \centering
\caption{Results for the Deezer dataset. Nodes in the visualization are colored based on their agreement values as illustrated in the color bar in the column of Densest Subgraph.
}
\label{tab:deezer}
\scalebox{0.79}{
    \begin{tabular}{c|c|c|c|c}\toprule
     Method & Densest Subgraph & AF ($\theta=0.5$) & $\methodlag$ ($\theta=0.5$) & $\method$ ($\theta=0.5$)    \\ \midrule
    Size & 1477 & 20 & 152 & 120 \\ 
    Density & 16.61 & 4.95 & 12.30 & 12.08 \\ 
    Agreement & $-0.23$ & 1.0 & 0.507 & 0.508 \\ \midrule
    Visualization & \multicolumn{1}{|m{3.3cm}|}{\resizebox{0.2\textwidth}{!}{%
        \includegraphics[width=\linewidth]{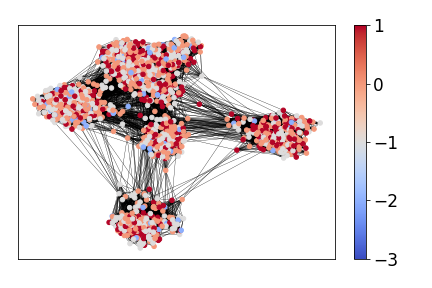}
        } }& \multicolumn{1}{|m{3.3cm}|}{ \resizebox{0.2\textwidth}{!}{%
        \includegraphics[width=\linewidth]{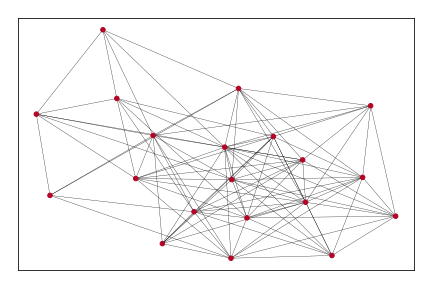}
        } }& \multicolumn{1}{|m{3.3cm}|}{ \resizebox{0.2\textwidth}{!}{%
        \includegraphics[width=\linewidth]{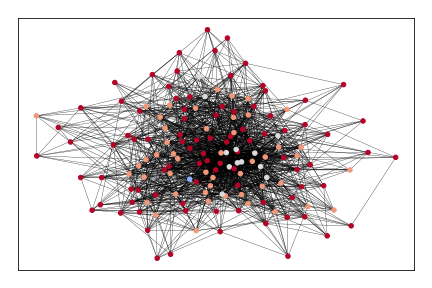}
        } }& \multicolumn{1}{|m{3.3cm}}{ \resizebox{0.2\textwidth}{!}{%
        \includegraphics[width=\linewidth]{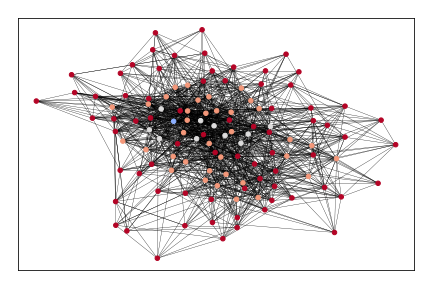}
        } }\\ \bottomrule
\end{tabular}
}
\end{table*}

\begin{figure}
    \centering
    \begin{subfigure}[]{.49\linewidth}
        \centering
        \includegraphics[width=\linewidth]{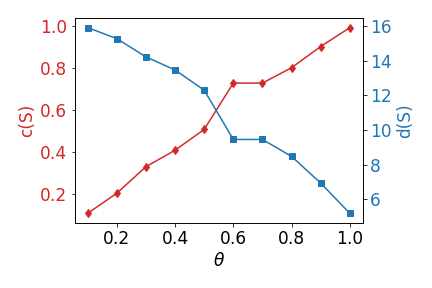}
        \vspace{-6mm}
        \caption{}
    \end{subfigure} \
    \begin{subfigure}[]{.49\linewidth}
        \centering
        \includegraphics[width=\linewidth]{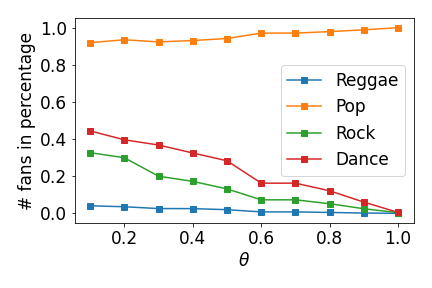}
        \vspace{-6mm}
        \caption{}
    \end{subfigure}
    \caption{Results for query vector $\bq_{\textsf{P}, \neg \textsf{RoDRe}}$. 
    (a) Trade-off between the the average agreement and the density of the output of \method as we range $\theta$. (b) The fraction of music fans liking each genre as a function of $\theta$.}
    \label{fig:tradeoff_deezer}
\end{figure}

\subsection{Results for Deezer dataset}
For this dataset, we conduct our performance test and case study simultaneously. Our formulation and algorithms are capable of deriving numerous insights from intricate datasets like Deezer, which combines elements of a music streaming platform with those of a social network. As an example, imagine a query vector $\bq_{\textsf{P}, \neg \textsf{RoDRe}}$ where $\mathsf{Pop}=+1$ and $\mathsf{Rock}$, $\mathsf{Dance}$, $\mathsf{Reggae}$ are set to $-1$, with all other music genres assigned a value of 0. This effectively sets up a distinction between \textsf{Pop} and a combined group of \textsf{Rock}, \textsf{Dance}, and \textsf{Reggae}, while remaining neutral towards the other genres. 

Table~\ref{tab:deezer} presents a summary of our findings, accompanied by visual representations of the subgraphs identified by the algorithms with $\theta=0.5$. 
As the LP baselines do not terminate in two hours, we omit them here.
The densest subgraph, not taking the agreement constraint into account, includes nodes of various similarities/dissimilarities, resulting in a subgraph with average degree $16.61$ and a total agreement sign with $\bq$ being negative. In contrast, the baseline AF extracts a subgraph comprising 20 nodes that achieves an upper bound of the agreement value of $1.0$ but seriously sacrifices the density. Our methods, \methodlag and \method, produce two separate subgraphs. These subgraphs primarily contain nodes that are similar, though they also include some nodes that are dissimilar. Nevertheless, they are much larger in size and significantly denser, serving as an effective middle ground that cannot be found by AF and the densest subgraph. Notably, the upper bound on the optimal value obtained by Proposition~\ref{prop:lagr} is $12.38$, meaning that both \methodlag and \method obtain near-optimal solutions.

We also study how the threshold $\theta$ affects the output of our methods for the same query vector $\bq_{\textsf{P}, \neg \textsf{RoDRe}}$. We present our findings for \method as we adjust $\theta$ from 0.1 to 1.0 in increments of 0.1. The results are depicted in Figure~\ref{fig:tradeoff_deezer}. With an increase in $\theta$, we note a compelling balance between agreement and density, aligning with our predictions, as shown in Figure~\ref{fig:tradeoff_deezer}(a). Figure~\ref{fig:tradeoff_deezer}(b) displays the percentage of individuals favoring each genre. The genres $\mathsf{Rock}$, $\mathsf{Dance}$, and $\mathsf{Reggae}$, which are less favored according to the query vector, start with small percentages at less than 0.5 and diminish to 0 as $\theta$ rises, as anticipated. Note that across a broad spectrum of lower $\theta$ values, the sum of proportions of individuals preferring each specific genre exceeds 1, due to the fact that a user can favor more than one music genres.

\subsection{Scalability}
So far, we have confirmed that our algorithms \methodlag and \method, as well as the simplest baseline AF, are scalable for moderately large graphs, while LP-S and LP-G have limited scalability. 
To further examine the scalability of our algorithms, we run them for the larger real-world graphs, with $\theta=0$. 
Table~\ref{tab:scale} summarizes the running time of the algorithms. 
As evident, both our methods are efficient and can scale to large networks. In particular, \method finishes in two minutes on the largest LiveJournal network, as it just performs a near-linear time procedure logarithmic times, as stated in Section~\ref{subsec:dual}. 
 \begin{table}[]
    \centering
    \caption{\label{tab:scale} Running time (sec) on large graphs.
    }
    \scalebox{0.7}{ 
    \begin{tabular}{c|c|c|c|c}\toprule
    Name  & AF & LP baselines& \begin{tabular}{c}\methodlag \end{tabular} & \begin{tabular}{c}\method \end{tabular} \\ \midrule
    Amazon & 3.77 & >3600 & 30.75 & 2.26 \\  
    DBLP-SNAP & 4.10 & >3600 & 28.12 & 2.11 \\ 
    YouTube & 21.62 & >3600 & 97.32 & 7.10 \\ 
    LiveJournal & 216.87 & >3600 & 720.36 & 125.48 \\
    \bottomrule
    \end{tabular}
    }
\end{table}

\section{Conclusions}
\label{sec:concl}
In this paper, we have introduced a novel formulation that aims to find a dense subgraph with opinion information aligned with a query vector. We studied the hardness of the problem and designed two algorithms \methodlag and \method. Our comprehensive analysis, supported by empirical evaluations on various real-world datasets, including those derived from Twitter on current issues like the Ukraine conflict and COVID-19 vaccine discussions, underscores the effectiveness of our algorithms. The ability to extract meaningful insights from opinion-laden graphs not only advances the field of network analysis but also offers valuable tools for understanding complex social phenomena and can enhance the performance of recommender systems. 
A promising area for future research is to develop efficient approximation algorithms that achieve the same approximation guarantee as D$k$S and to create algorithms tailored for temporal and multilayer networks~\cite{Kawase+23,petsinis2024finding}.

\clearpage

\section*{Ethical Considerations}
This paper involves the analysis of Twitter data; however, we have taken stringent measures to ensure the privacy and security of the information. The data used in our study has been fully anonymized, and no individual user information is made public or can be traced back to any person. We adhere strictly to data protection regulations to prevent any potential data leakage and uphold the confidentiality of the data subjects.

This paper explores new tools for dense subgraph discovery on social networks. While we focus on the algorithmic foundations,  the development and application of such  tools can have broader social impacts. For instance, while they can enhance decision-making in fields like advertising, they could also raise concerns regarding predictive policing or personal data profiling. It is crucial that the community and policy makers consider these potential impacts as we further develop and deploy these technologies.

\bibliographystyle{ACM-Reference-Format}
\balance
\bibliography{ref}

\clearpage
\setcounter{page}{1}
\appendix
\onecolumn

\section{Deferred Proofs}

Here we provide the proofs of the propositions stated earlier in the paper but deferred from the main text. 

\subsection{Proof of Proposition~\ref{prop:reduction}}\label{apdx:prop1}
\begin{proof}
We construct a polynomial-time reduction from Dam$k$S, which is known to be NP-hard~\cite{Khuller_Saha_09}, to \myproblem. 
Recall that in Dam$k$S, given an undirected graph $G=(V,E)$ and a positive integer $k$, we are asked to find $S\subseteq V$ that maximizes $d(S)$ subject to $|S|\leq k$. 
Let $(G=(V,E),k)$ be a given instance of Dam$k$S. 
We generate a gadget $(G'=(V',E'),(\bp_v)_{v\in V'},\bq,\theta)$ (i.e., an instance of \myproblem) as follows: 
The graph $G'$ is made from $G$ by adding $k$ singletons. Thus, $E'=E$ holds. 
For each $v\in V'$, set $\bp=(0)\in \bbR^1$ if $v$ comes from $V$ and $\bp=(+1)\in \bbR^1$ otherwise (i.e., if $v$ is a singleton added). 
Set the query opinion vector $\bq=(+1)\in \bbR^1$ and the threshold $\theta=1/2$. 
Then it is easy to see that there exists a feasible solution of Dam$k$S on $(G,k)$ with the objective value of $\textsf{val}$ if and only if there exists a feasible solution of \myproblem on $(G',(\bp_v)_{v\in V'},\bq,\theta)$ with the objective value of $\textsf{val}/2$. 
Therefore, we have the proposition. 
\end{proof}

\subsection{Proof of Proposition~\ref{prop:lagr}}\label{apdx:prop4}
\begin{proof} 
As $S_\mathrm{out}$ is an optimal solution to the Lagrangian relaxation with $\lambda^R$, we see that its objective value upper bounds the optimal value of \myproblem, i.e., 
$d(S_\mathrm{out})+\lambda^R(c(S_\mathrm{out})-\theta)\geq \mathrm{OPT}$,
which immediately obtains the statement. 
\end{proof}

\subsection{Proof of Proposition~\ref{prop:theory_dual}}\label{apdx:prop5}
\begin{proof}
From the behavior of Algorithm~\ref{alg:peel}, we see that $T$ is one of the candidate subsets of the output. Therefore, we have $d(S_\mathrm{out})\geq d(T)$, and hence in what follows, we lower bound $d(T)$. 
Note that $T$ is a subset that appears in the peeling procedure corresponding to $z_2^R$. 
By the definition of $\ell(T)$ in~\eqref{eq:load_subset}, we have that for any $v\in T$, 
$\deg_T(v) + z_2^R(c_v-\theta)\geq \ell(T)$, and thus 
$\deg_T(v)\geq \ell(T) -z_2^R(c_v-\theta)$. 
Summing up this over $v\in T$, we have 
\begin{align*}
\sum_{v\in T}\deg_T(v)\geq \ell(T)|T|-z_2^R\left(\sum_{v\in T}c_v-\theta|T|\right). 
\end{align*}
Dividing the both sides of the above inequality by $2|T|$, we have 
\begin{align*}
d(T)\geq \frac{\ell(T)}{2}-\frac{z_2^R}{2}\left(c(T)-\theta\right). 
\end{align*}
Recalling $\ell(T)\geq \mathrm{OPT}$, we can conclude the proof. 
\end{proof} 

\section{Supplementary Materials for Experiments}

Here we present additional materials and details related to the experiments described in the main text.

\subsection{Specifics of Twitter dataset}\label{appendix:twitter}

\noindent
\textbf{Collection.} One of the major contributions of our research to the broader community is the assembly of a Twitter dataset that we intend to release for public access. This dataset was compiled using twAwler~\cite{pratikakis2018twawler,sotiropoulos2019twittermancer} prior to Twitter's revision of its data collection policies. We concentrate on two contentious subjects that have garnered significant attention on Twitter: COVID-19 vaccination and the Russia--Ukraine conflict. We select an active user who expresses negative opinions towards both topics as the starting node, and expand a followship network in a forest-fire-like pattern. The crawling process results in a network consisting of 185\,533 users connected by 727\,138 undirected following relationships.

Regarding the vaccination discussion, we compile a list of relevant keywords and hashtags (`vaccine', `vaccination', `vaccinate', `vax', `vaccinated'), and classify a tweet as related if it was posted after 2019 and includes at least one of these keywords in its text or hashtags. For the topic of the Ukraine war, we use a set of keywords (`ukraine', `russia', `ukrainewar', `ukrainerussianwar', `war', `russiainvadedukraine'), with the requirement that the posts must date from after 2021. We exclude users from the network who have not shared tweets concerning either of these subjects. This filtering process yields a network comprising 3\,409 users and 11\,508 undirected connections, featuring over 80\,000 tweets related to these topics shared by the users.

\smallskip
\noindent
\textbf{Processing opinions.} We leverage the impressive developments in Large Language Models (LLMs) to obtain accurate estimates of the sentiment of the tweets. Similar to the observations in Zhang et al.~\cite{zhang-etal-2024-sentiment}, LLMs like GPT-3.5 outperform other models we tried, including BERT-based sentiment analysis models, when fine-tuning on this specific dataset is not provided. 
We evaluate the sentiments of these tweets regarding the topics using GPT-3.5.   Each tweet is assigned a score ranging from 0 to 10, where 5 indicates neutrality. In the context of the vaccination discussion, a score of 0 indicates complete opposition to vaccination, whereas a score of 10 signifies total support for it. Regarding the war topic, a score of 0 shows support for Russia (and by implication, opposition to Ukraine), while a score of 10 reflects support for Ukraine. Specifically, the prompts we use for querying tweet opinions are the following:

\begin{tcolorbox}[title=Tweet Rating Instructions, colback=white]
\begin{itemize}
\leftskip=-15pt
    \item Return only integer rate for the following tweet. The rate represents its opinion towards COVID vaccination with an integer between 0 to 10, with 10 being very positive and supportive of vaccination, 0 being very negative and skeptical about it, and 5 being completely neutral. COVID vaccination is also called vax.
    \item Return an integer in the range 0 to 10 (without additional comments, no letters at the sentence) for the following tweet. The rate represents its opinion towards a hypothetical war between Ukraine and Russia with an integer between 0 to 10, with 10 being totally supportive of Ukraine, 0 being totally supportive of Russia, and 5 being completely neutral.
\end{itemize}
\end{tcolorbox}

We convert the score to a range between $-1$ and $+1$ by subtracting five from each score and then dividing by five, aligning with the typical framework used in opinion dynamics research. Tweets deemed irrelevant by GPT to the discussed topic are excluded from analysis. Ultimately, we determine each user's overall opinion by calculating the average score of all relevant tweets they have shared.

\begin{table}[]
    \centering
    \caption{Performance of algorithms on the Twitter dataset with query $\bm{q}=(-1,1)$ and varying $\theta$. } \label{tab:comparetwitter} 
    \scalebox{0.9}{ 
    \begin{tabular}{c|c|c|c|c|c|c|c|c|c|c|c}\toprule
    & \multicolumn{2}{c|}{AF }  & \multicolumn{2}{c|}{LP-S } & \multicolumn{2}{c|}{LP-G } & \multicolumn{2}{c|}{\methodlag } & \multicolumn{2}{c|}{\method } & \\
    $\theta$  &  $d(S)$ &  $c(S)$  & $d(S)$ &  $c(S)$ &  $d(S)$ &  $c(S)$ &  $d(S)$ &  $c(S)$ &  $d(S)$ &  $c(S)$ & UB \\ \midrule
    0.2 & 9.29 & 0.48 & 18.68 &	 0.20 &	 18.11 & 0.21 & 18.68 & 0.20 & 17.52 & 0.35 & 19.08\\ 
    0.5 & 5.03 & 0.73 & 0.0 &	 1.80 &	 10.66&	 0.50 & 0.0 &	 1.80 & 14.83 & 0.50 & 16.90 \\ 
    0.8 & 1.50 & 1.00 & 0.0 &	 1.80 &	 4.78 &	 0.81 & 0.0 &	 1.80 &  7.65 & 0.80 & 13.00 \\ 
    1.1 & 0.83 & 1.27 & 0.0 &	 1.80 &	 1.94 &	 1.11 & 0.0 &	 1.80 &  2.82 & 1.11 & 9.10\\ \bottomrule
    \end{tabular}
    }
\end{table}

\subsection{Results for Twitter dataset}\label{appendix:experiments_twitter}

\begin{figure}
    \centering
    \includegraphics[width=0.3\linewidth]{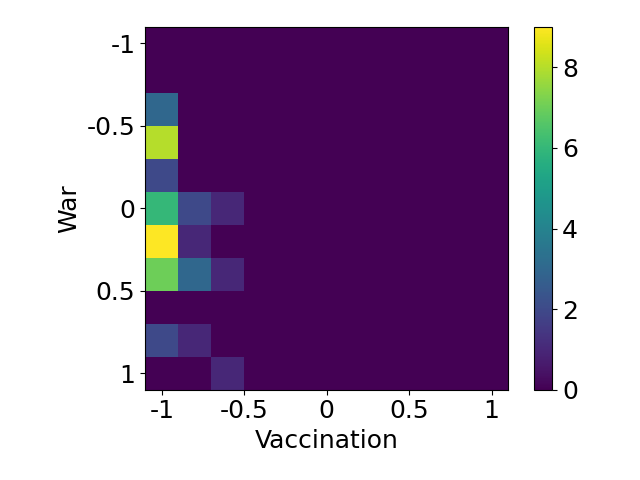}
    \caption{Histogram of opinions on (\textsf{Vax}, \textsf{Ukraine}) for the outputs of \method for $\theta=1.1$ with query $\bm{q}=(-1,1)$.}
    \label{fig:twit_theta_110}
\end{figure}

Table~\ref{tab:comparetwitter} shows the densities and average agreement values of the resulting subgraphs as we vary $\theta$ in $\{0.2,0.5,0.8,1.1\}$. 
We observe that \method again outperforms the baseline methods and \methodlag. 
Remarkably, \method outperforms 
AF not only when given the same $\theta$, but also when the empirical average agreement values are close. 
Figure~\ref{fig:twit_theta_110} depicts the histogram of opinions for the output of \method for $\theta=1.1$ with query $\bm{q}=(-1,1)$. 
Thanks to the larger $\theta=1.1$, the opinions tend to align with the query, unlike the output with $\theta=0.5$ captured in Figure~\ref{fig:twitter}(d).

\end{document}